\documentclass[aps,prl,twocolumn,superscriptaddress,secnumarabic]{revtex4-2}

\usepackage{amsmath}
\usepackage{graphicx}
\usepackage{braket}
\usepackage{soul}
\usepackage{siunitx}
\usepackage{balance}

\usepackage[usenames,dvipsnames]{color}
\graphicspath{{figures/}{}}

\begin{document}

\title{Photolithography-Only Fabrication of Transmons Using Double-Oblique Evaporation}

\author{K. Aoyanagi}
\affiliation{Department of Physics, Kyoto University, Kitashirakawa-Oiwakecho, Sakyo-ku, Kyoto 606-8502, Japan}

\author{S. Abe}
\affiliation{Department of Physics, Kyoto University, Kitashirakawa-Oiwakecho, Sakyo-ku, Kyoto 606-8502, Japan}

\author{S. Chen}
\affiliation{Department of Physics, Kyoto University, Kitashirakawa-Oiwakecho, Sakyo-ku, Kyoto 606-8502, Japan}

\author{T. Inada}
\affiliation{International Center for Elementary Particle Physics (ICEPP), The University of Tokyo, 7-3-1 Hongo, Bunkyo-ku, Tokyo 113-0033, Japan}

\author{C. Kawai}
\affiliation{Department of Physics, The University of Tokyo, 7-3-1 Hongo, Bunkyo-ku, Tokyo 113-0033, Japan}
\affiliation{QUP (WPI), KEK, OHO 1-1, TSUKUBA, IBARAKI 305-0801, JAPAN}

\author{Y. Mino}
\affiliation{International Center for Elementary Particle Physics (ICEPP), The University of Tokyo, 7-3-1 Hongo, Bunkyo-ku, Tokyo 113-0033, Japan}

\author{K. Nakamura}
\affiliation{Department of Physics, Kyoto University, Kitashirakawa-Oiwakecho, Sakyo-ku, Kyoto 606-8502, Japan}

\author{K. Nakazono}
\affiliation{Department of Physics, The University of Tokyo, 7-3-1 Hongo, Bunkyo-ku, Tokyo 113-0033, Japan}
\affiliation{QUP (WPI), KEK, OHO 1-1, TSUKUBA, IBARAKI 305-0801, JAPAN}

\author{T. Nitta}
\affiliation{QUP (WPI), KEK, OHO 1-1, TSUKUBA, IBARAKI 305-0801, JAPAN}

\author{K. Watanabe}
\affiliation{Department of Physics, The University of Tokyo, 7-3-1 Hongo, Bunkyo-ku, Tokyo 113-0033, Japan}
\affiliation{QUP (WPI), KEK, OHO 1-1, TSUKUBA, IBARAKI 305-0801, JAPAN}

\date{\today}

\begin{abstract}
We investigate a photolithography-only fabrication process for transmon Josephson junctions using a modified double-oblique evaporation geometry. Using a bilayer resist process and Al shadow evaporation, we fabricate junction structures and confirm by optical and scanning electron microscopy that the resulting narrowed crossing region reaches a geometrical area on the order of \(10^4~\mathrm{nm}^2\), which lies in the size range relevant to qubit junction fabrication. Room-temperature resistance screening shows that the junction resistance falls within the target range for the present transmon design over a usable process window and exhibits a clear design dependence. We further implement fabricated junctions in transmon devices and evaluate them in a three-dimensional Al cavity at \SI{20}{\milli\kelvin}, where we observe basic transmon qubit operation with \(f_{01}=\SI{4.865}{\giga\hertz}\), \(T_1 \sim \SI{9}{\micro\second}\), and \(T_2^* \sim \SI{0.4}{\micro\second}\). These results demonstrate the feasibility of realizing functional transmon devices in a photolithography-only process using double-oblique evaporation.
\end{abstract}

\maketitle

\section{Introduction}
Superconducting qubits, and in particular transmons, are among the leading platforms for quantum information processing because of their relatively simple circuit architecture, compatibility with microwave control, and scalability toward multi-qubit systems \cite{Koch}. Since the Josephson junction (JJ) provides the nonlinear inductance of the circuit and thereby determines key device parameters such as the qubit frequency and anharmonicity, establishing a stable and reproducible JJ fabrication process is a central issue in the development of superconducting quantum circuits. In standard transmon fabrication, large circuit features such as capacitor pads and control wiring are typically defined by photolithography, whereas the JJ is commonly fabricated by electron-beam lithography (EBL) combined with shadow evaporation \cite{Dolan,Niemeyer}. This division of roles originates from the fact that conventional optical lithography has generally lacked the resolution required to directly define the \(O(100~\mathrm{nm})\)-scale structures typically used for qubit JJs.

Although the EBL-based process has been highly successful and remains the dominant route for superconducting qubit fabrication, it also introduces several practical limitations. In particular, EBL is comparatively slow, costly, and less favorable for wafer-scale throughput and simplified process integration \cite{Osman}. These issues become increasingly important as superconducting quantum circuits move toward larger-scale integration and more routine device production. From this perspective, a fabrication route in which the JJ can also be realized using photolithography alone is attractive from the standpoint of cost, throughput, and process accessibility \cite{Monroe,Foroozani,Bal}. It is also compatible with the broader trend toward large-area fabrication \cite{VanDamme}.

Recent studies have begun to explore this direction. Monroe et al. demonstrated that transmon-compatible JJs can be fabricated using an all-optical direct-write process based on a Dolan-Niemeyer-bridge structure \cite{Monroe}. Their results showed that photolithography-only JJ fabrication is feasible when oxidation and evaporation conditions are appropriately engineered. In that work, the lithographically defined junction area was relatively large compared with conventional EBL-defined qubit junctions, and the junction properties were controlled primarily through process optimization. These results established the feasibility of photolithography-based transmon JJ fabrication. At the same time, obtaining smaller effective junction dimensions within a photolithographic process remains an important challenge, motivating the exploration of additional geometrical strategies.

A key challenge nevertheless remains for photolithography-only JJ fabrication. Since the minimum linewidth achievable by photolithography is generally much larger than that in standard EBL-based JJ processes, it is difficult to obtain a small junction area directly from the planar mask geometry alone. For this reason, control of the effective junction area through three-dimensional process design becomes essential. In oblique-angle evaporation schemes, including Manhattan-style and bridgeless shadow-evaporation processes \cite{Potts,Lecocq}, the final overlap area is governed not only by the nominal mask dimensions but also by the evaporation angle, resist profile, and relative deposition geometry \cite{Muthusubramanian}. As a result, the effective junction area can be made substantially smaller than the apparent linewidth defined by photolithography, providing a possible route to overcoming the intrinsic resolution limit of optical lithography without resorting to EBL.

In this work, we investigate a photolithography-only fabrication process for transmon Josephson junctions based on a Manhattan-style geometry. Beyond the standard oblique-angle evaporation scheme, we introduce a modified double-oblique evaporation method in which the deposition direction is tilted not only with respect to the substrate normal but also with respect to the resist-pattern orientation. The primary objective of this study is to establish the feasibility of transmon-compatible JJ fabrication within a photolithography-only process. At the same time, the proposed double-oblique geometry may provide an additional geometrical parameter relevant to the effective junction dimensions through shadowing. Using this process, we fabricate Josephson junctions, characterize them by room-temperature resistance measurements, and incorporate them into transmon devices measured in a three-dimensional setup.

\section{Fabrication}

\subsection{Double-Oblique Evaporation}

\begin{figure}[t]
  \centering
  \includegraphics[width=\columnwidth]{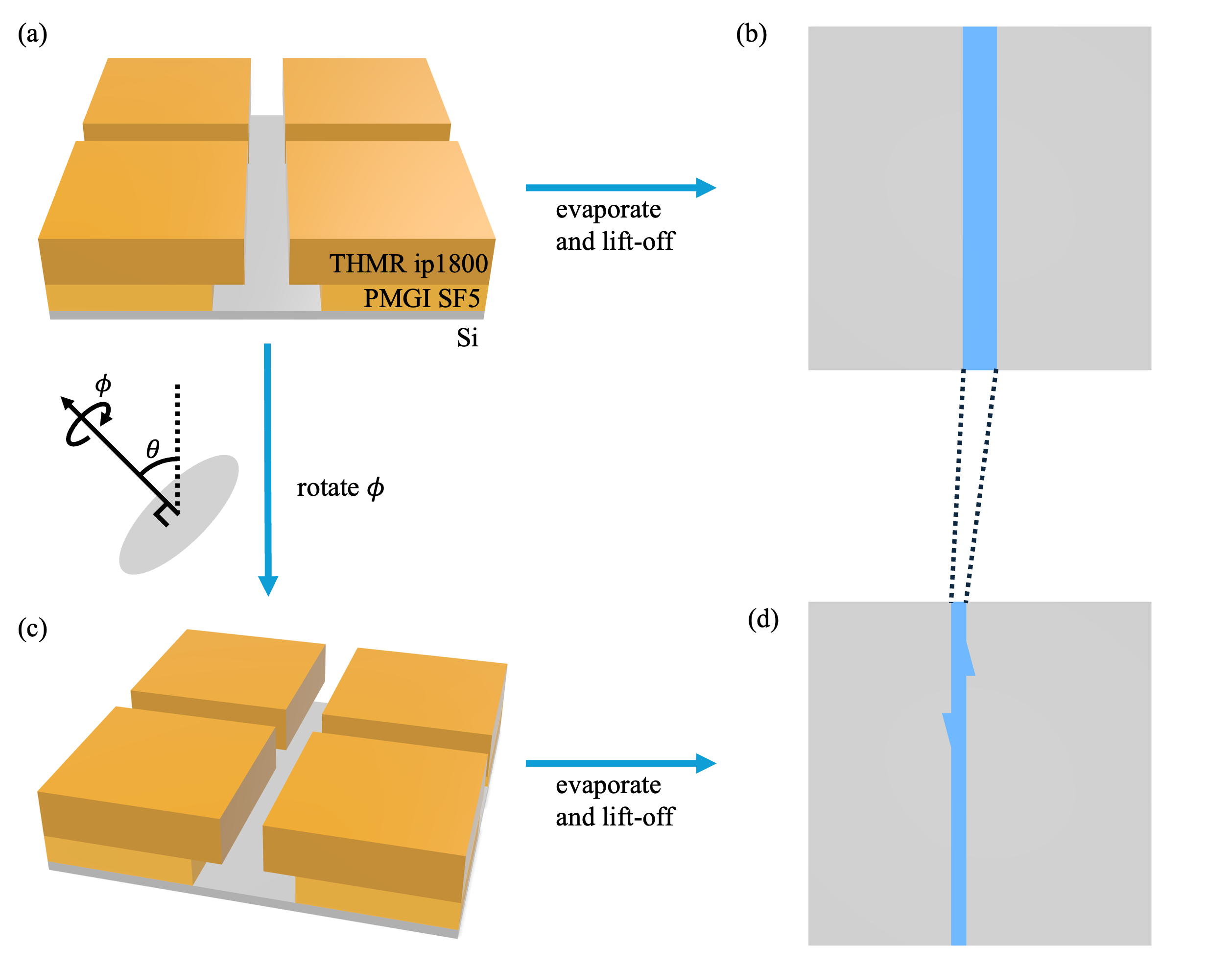}
  \caption{
    Schematic illustration of the double-oblique evaporation concept.
  (a) Perspective view of the bilayer resist opening in the conventional oblique-angle evaporation geometry.
  (b) Corresponding top view of the resulting evaporated region.
  (c) Perspective view of the modified double-oblique evaporation geometry, in which the deposition direction is tilted not only with respect to the substrate normal but also with respect to the resist-pattern direction.
  (d) Corresponding top view of the resulting evaporated region.
  }
  \label{fig:concept}
\end{figure}

Here, we define double-oblique evaporation as an evaporation geometry in which the deposition direction is oblique both to the substrate normal and to the resist-pattern orientation. We denote the deposition tilt angle from the substrate normal by $\theta$ and the in-plane angular offset between the deposition direction and the resist-pattern orientation by $\phi$. In this geometry, the resulting narrowed linewidth is determined not only by the nominal planar pattern, but also by the combined shadowing condition set by $\theta$, $\phi$, the linewidth of the resist opening, and the thickness of the upper resist layer.

Figure~\ref{fig:concept} schematically illustrates the geometrical concept of this method. In the conventional oblique-angle evaporation geometry, the deposition direction is tilted by $\theta$ with respect to the substrate normal while remaining aligned with the principal direction of the resist opening, i.e., $\phi = 0$, as shown in Fig.~\ref{fig:concept}(a,b). Under this condition, the linewidth of the deposited feature is mainly determined by the photolithographically defined opening width together with the standard shadowing profile of the bilayer resist.

In the present method, the sample is additionally rotated in-plane so that $\phi \neq 0$, as shown in Fig.~\ref{fig:concept}(c,d). The deposition direction is therefore oblique not only to the substrate normal but also to the resist-pattern direction. This double-oblique geometry modifies the shadowing condition inside the opening and laterally shifts the shadow boundary, resulting in a narrower deposited line than in the conventional case. Consequently, the in-plane angular offset $\phi$, together with $\theta$, the upper-resist thickness, and the nominal resist linewidth, constitutes a geometrical parameter set that influences the effective junction linewidth below the apparent photolithographic linewidth.

The dependence of the idealized narrowed linewidth on the geometrical parameters \(\theta\), \(\phi\), the upper-resist thickness, and the nominal opening width is summarized in Appendix~A.

\subsection{Fabrication and structural characterization}

\begin{figure}[t]
  \centering
  \includegraphics[width=\columnwidth]{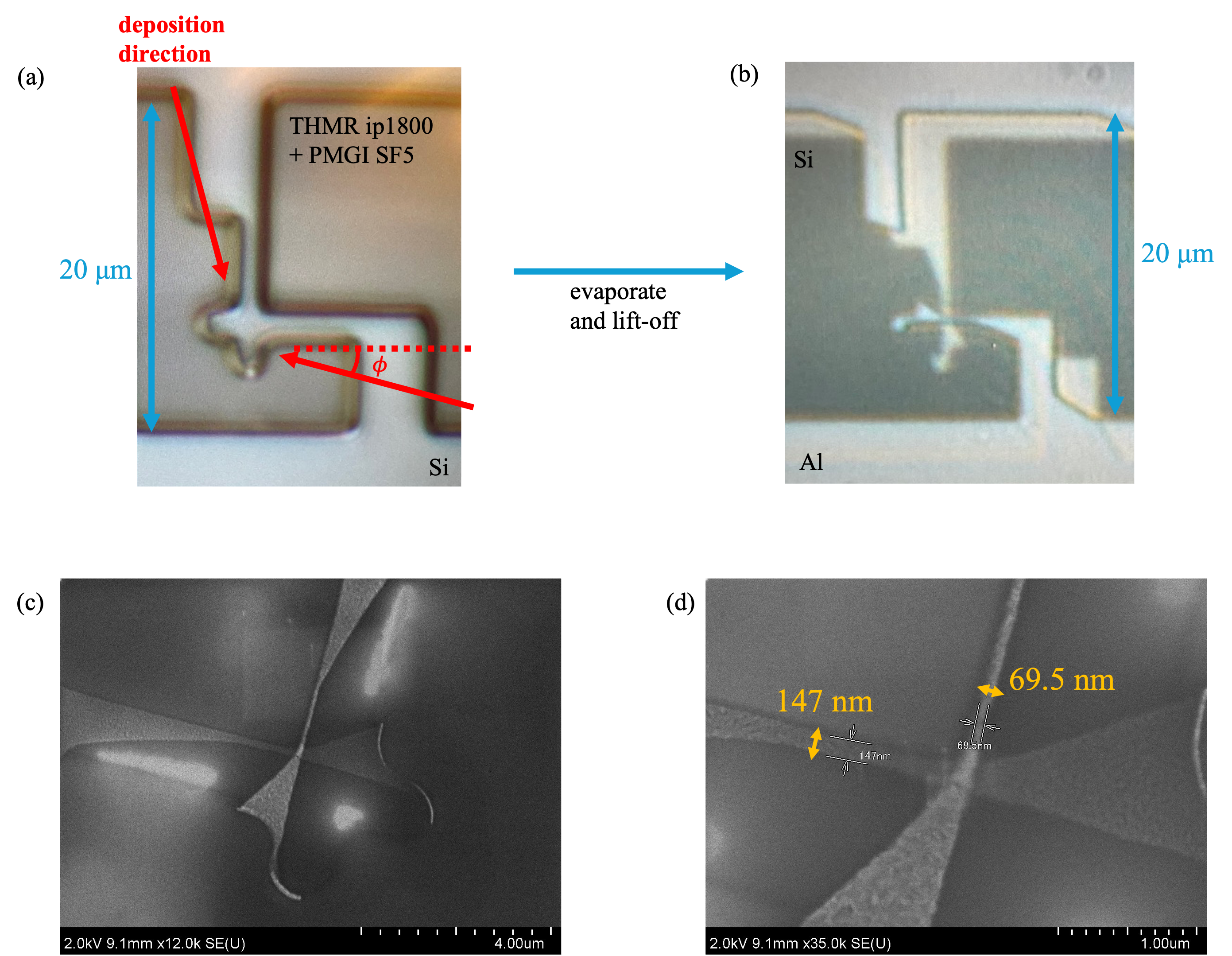}
  \caption{
  Optical and SEM images of the fabricated double-oblique structure.
  (a) Optical micrograph of the bilayer resist pattern before evaporation, with the deposition direction indicated by the red arrows.
  (b) Optical micrograph of the corresponding structure after evaporation and lift-off.
  (c) Low-magnification SEM image of the fabricated junction region.
  (d) High-magnification SEM image of the narrowed line segment.
  }
  \label{fig:real}
\end{figure}

The devices were fabricated on Si substrates using a photolithography-only process combined with the double-oblique evaporation scheme introduced above. Before resist coating, the substrates were cleaned in 1:8 diluted buffered hydrofluoric acid (BHF) for 1~min to remove the native oxide. A bilayer resist stack was then prepared using PMGI SF5 as the underlayer and THMR iP1800 as the top imaging resist, with target thicknesses of approximately 200~nm and 1.8~\si{\micro\meter}, respectively. The junction patterns were defined by direct-write photolithography using a Heidelberg Instruments MLA150 maskless aligner operated with a 375 nm optical beam, and were developed in SD-1 developer, yielding the bilayer resist profile required for shadow evaporation. After development, a descum step was performed by oxygen plasma ashing using a Samco RIE-10NR-KF system operated for 30~s at 50~W with an O$_2$ flow rate of 100~sccm.

Al deposition was performed in a Plassys Bestek / MEB 550S2-HV evaporation system at a deposition rate of 0.2 nm/s. In all devices, the deposition tilt angle was fixed at \(\theta=\ang{60}\). For the structural characterization and room-temperature screening samples, the double-oblique geometry was realized with an in-plane angular offset of \(\phi=\ang{15}\). In the present design, the evaporation direction was intentionally chosen to point toward the outer side of the corner forming the junction crossing, as illustrated in Fig.~\ref{fig:real}(a). This configuration modifies the shadowing condition at the corner and narrows the local linewidth at the junction region. For the qubit device used in the millikelvin measurements, the same deposition tilt angle \(\theta=\ang{60}\) was used, while the in-plane angular offset was changed to \(\phi=\ang{25}\). The first and second Al evaporations had nominal thicknesses of 40~nm and 100~nm, respectively. The AlO\(_x\) tunnel barrier was formed \textit{in situ} between the two depositions by static oxidation in O\(_2\) at 0.3~Torr for 300~s in the Plassys chamber. After the second evaporation, lift-off was performed in N-methyl-2-pyrrolidone (NMP).

To confirm that the intended geometrical effect is realized in the actual process, we examined the fabricated structures by optical microscopy, as shown in Fig.~\ref{fig:real}. Figure~\ref{fig:real}(a) shows an optical micrograph of the bilayer resist pattern before metal deposition, together with the deposition direction used in the experiment; the corresponding in-plane angular offset for this structure is \(\phi=\ang{15}\). After evaporation and lift-off, the corresponding deposited structure is shown in Fig.~\ref{fig:real}(b). The deposited feature follows the expected double-oblique geometry and exhibits narrowed line segments at the intended junction location.

Further structural confirmation is obtained from the SEM images in Fig.~\ref{fig:real}(c,d). The low-magnification SEM image in Fig.~\ref{fig:real}(c) shows the overall geometry of the fabricated junction region, while the magnified image in Fig.~\ref{fig:real}(d) resolves the narrowed crossing region in greater detail. From this image, the linewidths of the two narrowed electrode segments are measured to be approximately 150~nm and 70~nm, respectively. Their intersection therefore defines a geometrical junction area of approximately \(10^4~\mathrm{nm}^2\), indicating that the present process can access the size range relevant to conventional qubit junction fabrication. The observed asymmetry between the two linewidths is attributed to the fact that the Al deposited in the first evaporation remains on the top surface and sidewall of the resist, thereby modifying the effective shadowing geometry for the second evaporation. This asymmetry is therefore considered to originate from the present evaporation geometry and resist profile, and may be mitigated by further design optimization. These observations support the expected effect of the double-oblique geometry, namely that the local junction dimensions can be reduced well below the apparent linewidth defined by the photolithographic opening.

\section{Measurement Results}

\subsection{Room-Temperature Characterization}

\begin{figure}[t]
\includegraphics[width=\columnwidth]{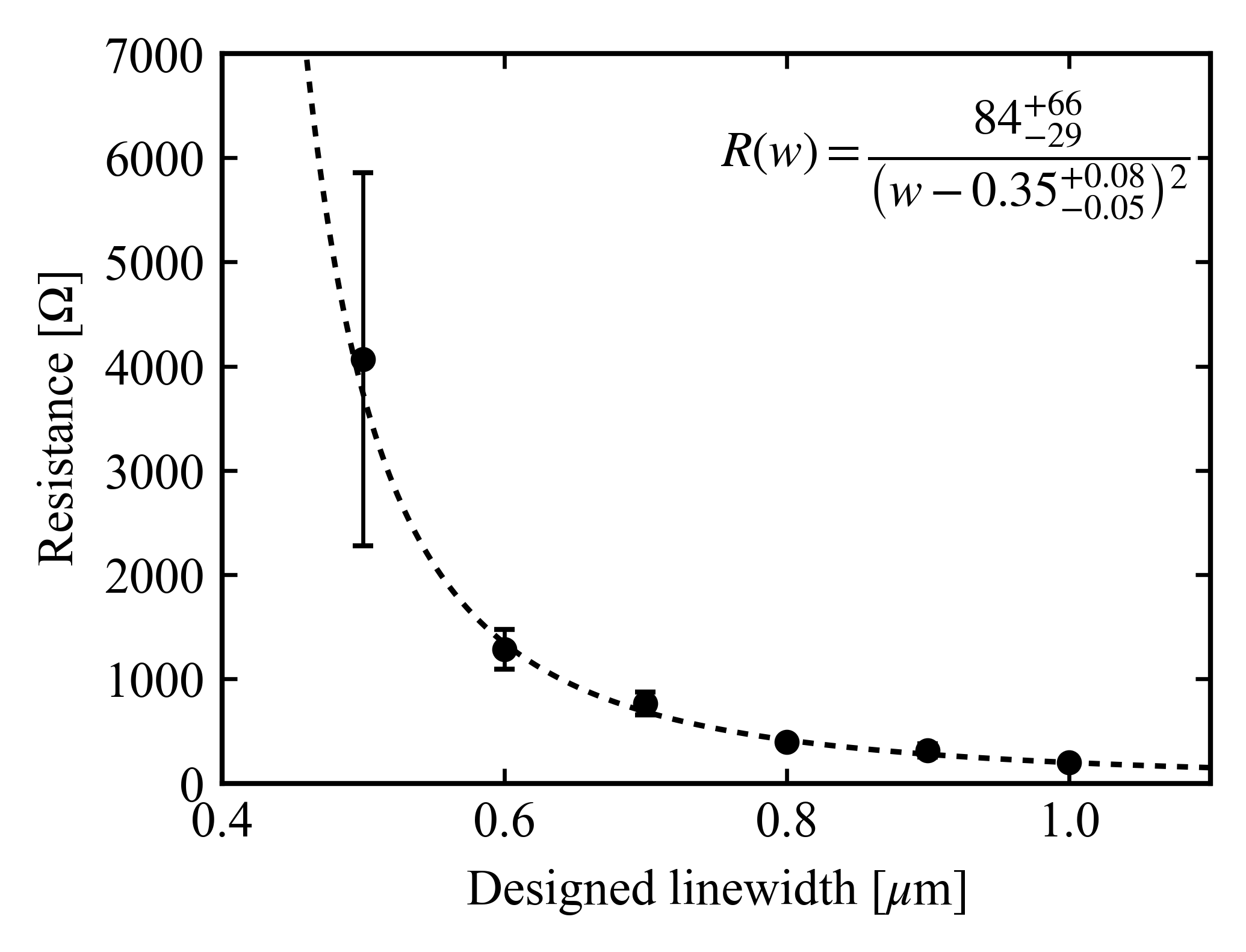}
\caption{
Room-temperature resistance as a function of designed linewidth.
The plotted values were obtained after subtraction of the constant offset associated with the two-terminal measurement.
Error bars indicate the standard deviation of the corrected resistance for devices that were not classified as shorts ($R \le 300~\si{\ohm}$) or open circuits ($R \ge 50~\si{\kilo\ohm}$).
The dashed line is a fit to
$R(w)=84^{+66}_{-29}/(w-0.35^{+0.08}_{-0.05})^2$,
where $R$ is in \si{\ohm} and $w$ is in \si{\micro\meter}.
}
\label{fig:rt_resistance}
\end{figure}

\begin{table}[t]
\caption{
Room-temperature screening results for junction test structures with different designed linewidths.
For each design, 27 devices were measured.
Devices with $R \le 300~\si{\ohm}$ were classified as shorts, and those with
$R \ge 50~\si{\kilo\ohm}$ were classified as open circuits and excluded from the resistance statistics.
The yield is given as the number of devices that were not excluded by these criteria.
The resistance values listed in the table are corrected values obtained after subtraction of the constant offset associated with the two-terminal measurement.
}
\label{tab:screening}
\begin{ruledtabular}
\begin{tabular}{cccc}
Designed width [\si{\micro\meter}] & Yield & Mean $R_{\mathrm{corr}}$ [\si{\ohm}] & Std.\ dev.\ [\si{\ohm}] \\
\hline
0.5 & 12/27 & 4066 & 1790 \\
0.6 & 23/27 & 1285 & 188  \\
0.7 & 26/27 & 764  & 112  \\
0.8 & 27/27 & 395  & 46   \\
0.9 & 25/27 & 314  & 64   \\
1.0 & 24/27 & 197  & 22   \\
\end{tabular}
\end{ruledtabular}
\end{table}

Room-temperature resistance measurements were performed on the screening samples fabricated at \(\theta=\ang{60}\) and \(\phi=\ang{15}\) in order to evaluate the design dependence and sample-to-sample variation of the junction resistance. For each design, 27 devices were measured. The measured resistance values were used to examine whether the design-dependent central values are consistent with the effective junction area inferred from the optical and SEM observations, and to quantify the sample-to-sample variation as an indicator of process reliability. Since the room-temperature normal-state resistance is directly related to the junction critical current, it provides a useful metric for process control and device screening \cite{Monroe,Foroozani,Ambegaokar}.

Figure~\ref{fig:rt_resistance} shows the corrected room-temperature resistance as a function of the designed linewidth. The resistance decreases monotonically with increasing linewidth, consistent with the increase in the effective junction area. To compare the junction resistance more directly, a constant offset associated with the two-terminal measurement was subtracted from the measured mean resistance for each design, and the resulting corrected values are summarized in Table~\ref{tab:screening} together with the yield and the standard deviation.

As shown in Table~\ref{tab:screening}, the 0.5--0.6~\si{\micro\meter} designs fall within the target resistance range for the present transmon design, approximately \SI{1}{\kilo\ohm} to \SI{10}{\kilo\ohm}. This indicates that the present geometry can bring the effective junction dimensions into the range relevant to transmon fabrication even within a photolithographic linewidth window. Among the \(\phi=\ang{15}\) screening samples, the 0.6~\si{\micro\meter} geometry exhibits a substantially higher yield (23/27) than the 0.5~\si{\micro\meter} geometry (12/27), while maintaining a corrected mean resistance of approximately \SI{1.3}{\kilo\ohm}. The 0.5~\si{\micro\meter} design gives a corrected mean resistance of approximately \SI{4.1}{\kilo\ohm}, but with a significantly reduced yield. The reduced yield at 0.5~\si{\micro\meter} may be related to the fact that this design approaches the practical linewidth limit of the present photolithography process. Taken together, these results indicate that the present double-oblique process can access the target junction resistance without requiring the smallest nominal photolithographic linewidth.

\subsection{Millikelvin Characterization}

\begin{figure}[b]
  \centering
  \includegraphics[width=\columnwidth]{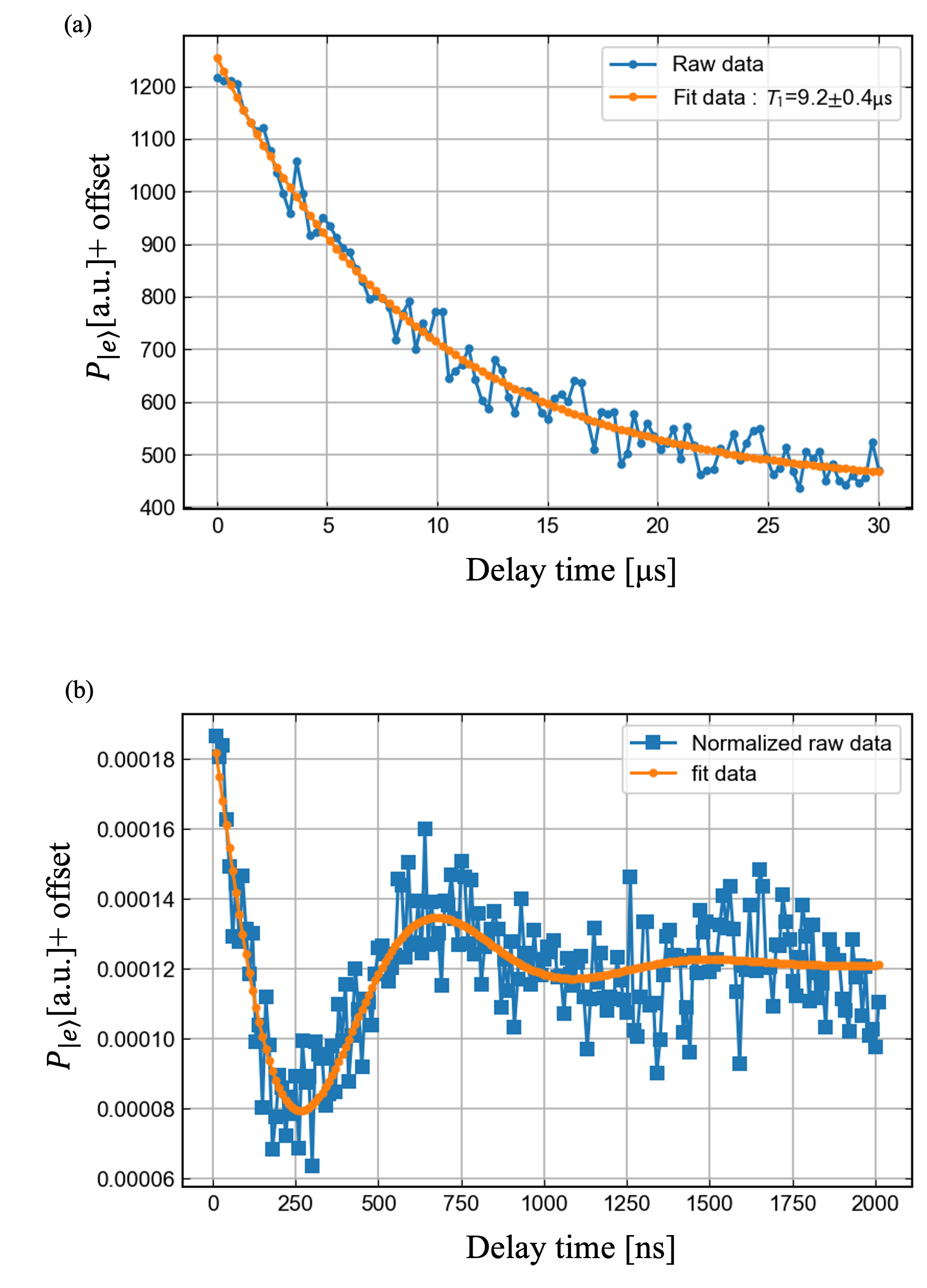}
  \caption{
    Low-temperature characterization of the fabricated transmon qubit in a three-dimensional Al cavity at \SI{20}{\milli\kelvin}.
    (a) Energy-relaxation measurement with a single-exponential fit, yielding \(T_1 = \SI{9.2 \pm 0.4}{\micro\second}\).
    (b) Ramsey measurement with a fit to a decaying oscillation, from which a dephasing time of approximately \(T_2^* \sim \SI{0.4}{\micro\second}\) is obtained.
  }
  \label{fig:T1T2}
\end{figure}

To evaluate the low-temperature properties of the fabricated devices, we implemented the qubit in a three-dimensional architecture using an Al resonant cavity and performed microwave measurements in a dilution refrigerator at a base temperature of approximately \SI{20}{\milli\kelvin} \cite{Paik}. The cavity resonance frequency was \SI{7.17}{\giga\hertz}, and the loaded quality factor was approximately \(Q \sim 4\times10^4\). For the measurements presented below, we used a qubit fabricated with \(\theta=\ang{60}\) and \(\phi=\ang{25}\), based on the 0.5~\si{\micro\meter} design and having a room-temperature resistance of approximately \SI{10}{\kilo\ohm}. The transmon capacitor pads were defined simultaneously with the Josephson junction electrodes during the Al evaporation process. The low-temperature measurements presented below demonstrate that a device fabricated under the \(\phi=\ang{25}\) condition exhibits the basic properties of a transmon qubit.

Figure~\ref{fig:T1T2}(a) shows the energy-relaxation measurement of the fabricated qubit. Fitting the decay curve with a single exponential yields a relaxation time of \(T_1 = \SI{9.2 \pm 0.4}{\micro\second}\). The qubit transition frequency was measured to be \(f_{01}=\SI{4.865}{\giga\hertz}\). Figure~\ref{fig:T1T2}(b) shows the Ramsey measurement, from which we estimate a dephasing time of approximately \(T_2^* \sim \SI{0.4}{\micro\second}\) by fitting the data to a decaying oscillation.

Although the coherence times are limited, the observation of a well-defined qubit transition together with finite \(T_1\) and \(T_2^*\) demonstrates that the fabricated device operates as a superconducting qubit. These results therefore establish the feasibility of realizing functional transmon devices using the present photolithography-only double-oblique evaporation process.

\section{Discussion}

The present results should be interpreted primarily as a feasibility study of photolithography-only transmon Josephson junction fabrication using double-oblique evaporation. Room-temperature screening was performed for samples fabricated at \(\phi=\ang{15}\), whereas the low-temperature qubit characterization was carried out using a device fabricated at \(\phi=\ang{25}\). In the present work, the room-temperature data provide an evaluation of design dependence and process reliability, while the low-temperature data establish that a device fabricated under a double-oblique condition can exhibit the basic properties of a transmon qubit. Direct comparison between room-temperature screening and qubit performance under identical \(\phi\) conditions will therefore be necessary in future work.

Another limitation of the present study is that the low-temperature characterization is based on a single device rather than on a statistically meaningful set of qubits. The measured values, \(f_{01}=4.865~\mathrm{GHz}\), \(T_1 \sim 9~\mu\mathrm{s}\), and \(T_2^* \sim 0.4~\mu\mathrm{s}\), are sufficient to establish qubit operation, but they do not define an optimized performance level for the present process. In sub-micron junction-based superconducting qubits, dielectric and capacitive losses associated with the junction electrodes and their surrounding interfaces can contribute significantly to energy relaxation \cite{Dunsworth}. The limited coherence observed here is therefore compatible with the simplified fabrication route adopted in the present study, although further optimization of the junction environment, materials processing, and device geometry will be required to identify the dominant loss mechanisms. Low-temperature characterization of multiple devices fabricated under the same screening condition will also be necessary to establish the range of qubit performance attainable with the present process.

The room-temperature screening and low-temperature qubit measurements together suggest that double-oblique evaporation can be incorporated into a simplified shadow-evaporation-based process while remaining compatible with basic transmon operation. At the same time, the present work should not be regarded as a demonstration of a scalable or foundry-compatible qubit fabrication process in the same sense as recent large-area optical-lithography-based approaches. Rather, the significance of the present method lies in providing a complementary route toward reduced dependence on conventional EBL-based Josephson-junction fabrication \cite{Osman}. A systematic comparison of the double-oblique parameters, in particular the dependence on \(\phi\), together with a control experiment using conventional oblique-angle evaporation under otherwise similar conditions, will be necessary to clarify how the in-plane angular offset modifies the effective overlap geometry, how this is reflected in the room-temperature resistance distribution, and whether the same design strategy can be used to improve the balance among nominal linewidth, target junction resistance, and fabrication yield.

\section{Conclusion}

We investigated a photolithography-only process for transmon Josephson junction fabrication using double-oblique evaporation. Structural characterization showed that the fabricated junctions exhibit narrowed crossing regions with geometrical dimensions in the size range relevant to transmon fabrication, and room-temperature measurements showed that the target junction resistance for the present transmon design can be reached within a usable process window.

We further implemented fabricated junctions in transmon devices and evaluated their performance in a three-dimensional Al cavity at \SI{20}{\milli\kelvin}. For the low-temperature measurements, we used a device fabricated under a different double-oblique condition, namely \(\phi=\ang{25}\). The measured qubit parameters, \(f_{01}=4.865~\mathrm{GHz}\), \(T_1 \sim 9~\mu\mathrm{s}\), and \(T_2^* \sim 0.4~\mu\mathrm{s}\), confirm basic qubit operation. These results suggest that double-oblique evaporation provides a viable route toward simplified transmon Josephson junction fabrication in a photolithography-only process.

\section{Acknowledgments}

The authors acknowledge support from the International Center for Quantum-field Measurement Systems for Studies of the Universe and Particles (QUP), a WPI Research Center Initiative at KEK, and the International Center for Elementary Particle Physics (ICEPP), The University of Tokyo.
The cryogenic measurements were carried out using facilities at QUP and the joint-use facilities of the Millikelvin Quantum Platform at the Cryogenic Research Center, The University of Tokyo.
This work was also supported by the External Use Program of the OIST Core Facilities, Okinawa Institute of Science and Technology Graduate University.
This work was supported in part by JSPS KAKENHI Grant Numbers JP23H01182, JP23H04864, JP24H00689, JP25H00638, JP25KJ0847, and JP26KJ0842; JST PRESTO Grant Number JPMJPR23F7; JST SPRING Grant Number JPMJSP2108; and IBM--UTokyo Sponsored Research.




\appendix

\setcounter{figure}{0}
\setcounter{equation}{0}
\setcounter{table}{0}

\renewcommand{\thefigure}{A\arabic{figure}}
\renewcommand{\theequation}{A\arabic{equation}}
\renewcommand{\thetable}{A\arabic{table}}

\section*{Appendix A: Idealized geometrical dependence of the narrowed linewidth}

For reference, we summarize an idealized geometrical model for the linewidth reduction in the double-oblique evaporation scheme. The purpose of this appendix is to clarify how the narrowed linewidth depends on the geometrical parameters in the absence of non-ideal effects such as metal accumulation on the resist top surface or sidewall, re-emission, or wraparound deposition. The model should therefore be regarded as an idealized design guide rather than a quantitative description of the full fabricated structure.

\begin{figure}[t]
  \centering
  \hspace*{0.06\columnwidth}
  \includegraphics[width=0.7\columnwidth]{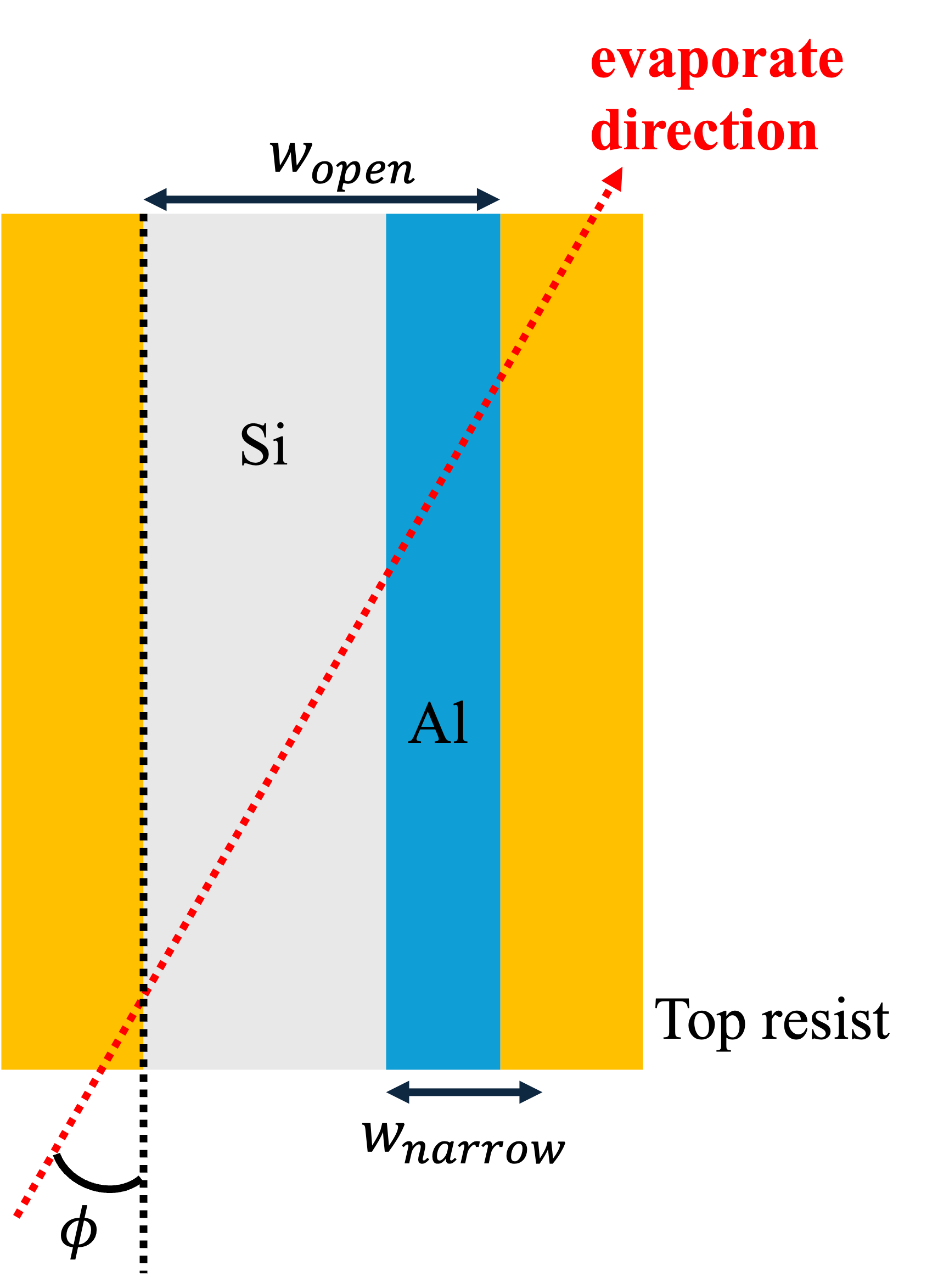}
  \caption{
    Top-view definition of the idealized narrowed linewidth in double-oblique evaporation. The nominal opening width is denoted by \(w_{\mathrm{open}}\), the resulting deposited linewidth by \(w_{\mathrm{narrow}}\), and the in-plane angular offset between the evaporation direction and the resist-pattern orientation by \(\phi\).
  }
  \label{fig:appendix_topview}
\end{figure}

\begin{figure}[tb]
  \centering
  \includegraphics[width=\columnwidth]{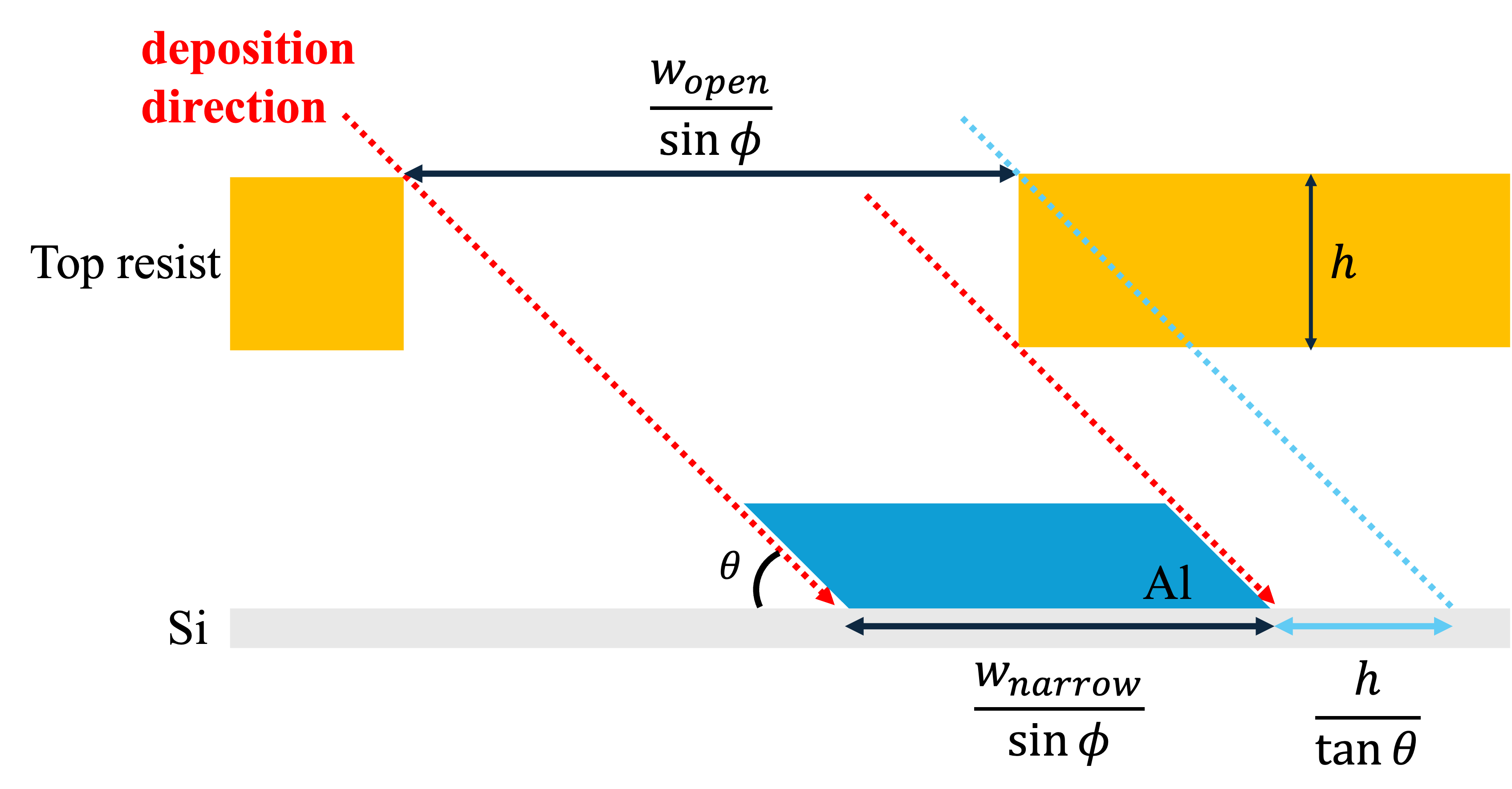}
  \caption{
    Idealized projected side-view geometry for the double-oblique evaporation scheme. The opening length projected along the evaporation direction is \(w_{\mathrm{open}}/\sin\phi\), and the projected narrowed deposited length is \(w_{\mathrm{narrow}}/\sin\phi\). The upper resist has thickness \(h\), and the deposition tilt angle from the substrate normal is \(\theta\). In this idealized geometry, the shadowing by the upper resist shortens the available deposited length by \(h/\tan\theta\).
  }
  \label{fig:appendix_sideview}
\end{figure}

We consider a resist opening with nominal width \(w_{\mathrm{open}}\), upper-resist thickness \(h\), deposition tilt angle \(\theta\) from the substrate normal, and in-plane angular offset \(\phi\) between the evaporation direction and the resist-pattern orientation. Figure~\ref{fig:appendix_topview} shows the top-view definition of the narrowed linewidth \(w_{\mathrm{narrow}}\), and Fig.~\ref{fig:appendix_sideview} shows the corresponding projected side-view geometry.

In the top view, the opening width projected along the evaporation direction is
\begin{equation}
L_{\mathrm{open}}=\frac{w_{\mathrm{open}}}{\sin\phi},
\label{eq:appendix_Lopen}
\end{equation}
and the corresponding projected length of the narrowed deposited region is
\begin{equation}
L_{\mathrm{narrow}}=\frac{w_{\mathrm{narrow}}}{\sin\phi}.
\label{eq:appendix_Lnarrow}
\end{equation}
In the idealized projected side view, the shadow cast by the upper resist shortens the available deposited length on the substrate by
\begin{equation}
\Delta L=\frac{h}{\tan\theta}.
\label{eq:appendix_dL}
\end{equation}
The projected narrowed length is therefore given by
\begin{equation}
L_{\mathrm{narrow}}=L_{\mathrm{open}}-\Delta L.
\label{eq:appendix_relation}
\end{equation}
Substituting Eqs.~(\ref{eq:appendix_Lopen})--(\ref{eq:appendix_dL}) into Eq.~(\ref{eq:appendix_relation}), we obtain
\begin{equation}
\frac{w_{\mathrm{narrow}}}{\sin\phi}
=
\frac{w_{\mathrm{open}}}{\sin\phi}
-
\frac{h}{\tan\theta},
\end{equation}
which yields
\begin{equation}
w_{\mathrm{narrow}}
=
w_{\mathrm{open}}
-
\frac{h\sin\phi}{\tan\theta}.
\label{eq:appendix_wnarrow}
\end{equation}

Equation~(\ref{eq:appendix_wnarrow}) shows that, within this idealized picture, the narrowed linewidth is determined by the nominal opening width \(w_{\mathrm{open}}\), the upper-resist thickness \(h\), the deposition tilt angle \(\theta\), and the in-plane angular offset \(\phi\). In particular, increasing \(h\) or \(\phi\), or decreasing \(\theta\), enhances the geometrical narrowing. In the actual fabricated structures, however, the first evaporated metal can remain on the resist top surface and sidewall, thereby modifying the effective shadowing geometry for subsequent deposition. Deviations from Eq.~(\ref{eq:appendix_wnarrow}) are therefore expected in the experiment, especially when non-ideal resist profiles and metal accumulation become significant.

\balance
\bibliography{paper}

\end{document}